\documentstyle[12pt]{article}
\renewcommand{\thefootnote}{\fnsymbol{footnote}}

\begin{document}
\begin{titlepage}
\rightline{\vbox{\halign{&#\hfil\cr
&ANL-HEP-PR-94-41\cr
&IFT-12-94\cr
&July 1994\cr}}}
\vspace{1in}
\begin{center}

{\Large\bf
Renormalization Scheme Dependence and \\
the Problem of Theoretical Uncertainties in\\
Next-Next-to-Leading Order QCD Predictions}
\footnote{Work supported by KBN grant no. 202739101,
 and the U.S. Department of
Energy, Division of High Energy Physics, Contract W-31-109-ENG-38.}
\medskip

\normalsize Piotr A. R\c{a}czka
\footnote{Permanent address: Institute of Theoretical Physics,
 Department of Physics, Warsaw University, ul.Ho\.{z}a~69,
 00-681 Warsaw, Poland. Internet: praczka@fuw.edu.pl}
\\ \smallskip
High Energy Physics Division,\\
 Argonne National Laboratory,\\
 Argonne, IL 60439\\
\end{center}

\begin{abstract}
 Renormalization scheme uncertainties in the
 next-next-to-leading order QCD predictions are discussed.
 To obtain an estimate of these uncertainties
 it is proposed to compare
 predictions in all schemes that do not have unnaturally
 large expansion coefficients. A concrete prescription for
  eliminating the unnatural schemes is given, based on the
 requirement that large cancellations in the expression
 for the characteristic renormalization scheme invariant
 should be avoided. As an example the QCD corrections to
 the Bjorken sum rule  are considered. The
 importance of the next-next-to-leading order corrections
 for a proper evaluation of perturbative QCD predictions
 is emphasized.
\end{abstract}

\renewcommand{\thefootnote}{\arabic{footnote}}
\end{titlepage}

 In the process of evaluation of perturbative QCD predictions one inevitably
 encounters the problem of renormalization scheme (RS) dependence. As is
 well known, despite the fact that final predictions of a
 quantum field theory for physical quantities
 should be independent of choice of
 the renormalization scheme, the results obtained with the truncated
 perturbation series may be numerically different in different schemes.
 Although the difference is formally of higher order in the coupling
 constant, numerically it may be significant in the
 phenomenological analysis.
 This effect complicates somewhat comparison of the
 theoretical predictions
 with the experimental data. There has been a lively discussion
 on the problem of a proper choice of the RS
 \cite{GrunFAC,PMS,CelSiv,DharGuptaMOM,BLM,DukeRoberts}.

 However, the problem of a proper choice of the RS is only one of
 the aspects of the RS dependence effect. In fact, the RS dependence
 of perturbative predictions, instead of being an annoying complication,
 may also become a source of an important information. The idea is to
 estimate the strength of the RS dependence,
 and to use it  as a measure of reliability of perturbation expansion.
 Indeed, we expect the perturbative expression
 to be less reliable when the expansion coefficients are large and/or
 when the coupling constant is not very small. These are precisely
 the conditions under which we expect also a strong RS dependence of
 perturbative predictions.

 A procedure that is frequently used to estimate the strength of the
 RS dependence consists of calculating the variation of predictions
 when the renormalization scale is changed in some reasonable range.
 Unfortunately, in the next-next-to-leading order (NNLO) approximation
 the freedom of choice of RS is parametrized by {\em two} independent
 parameters, so that varying the renormalization scale we scan only
 a small subset of all the available schemes. Also, as has been
 discussed for example in \cite{R92}, the concept of a ``reasonable''
 renormalization scale is not well defined.

 A more general approach was
 formulated in \cite{R92}, in which full freedom of choice of the RS
 is taken into account. This approach is based on the
 observation that in NNLO
 one may select acceptable schemes using the expression for a RS
 invariant combination of the expansion coefficients. In this note
 the ideas presented in \cite{R92} are further developed. A precise
 mathematical formulation of the constraint eliminating the
 unnatural
 schemes is given, which allows for a quantitative estimate of the
 strength of RS dependence for any physical observable for which
 the NNLO corrections are known. The QCD corrections to the Bjorken
 sum rule for polarized structure functions are discussed as an example.
 The results obtained for other observables are also briefly
 summarized \cite{conf}.

 Let us consider a NNLO expression for a physical quantity $R$
 depending
 on a single energy variable $P$:
\begin{equation}
R^{(2)}(P) = a(\mu)[1+r_{1}(\mu/P)a(\mu)+r_{2}(\mu/P)a^{2}(\mu)],
\label{R}
\end{equation}
 where $\mu$  dentes the renormalization scale,
 $a(\mu)=g^{2}(\mu)/(4 \pi^{2})$
 denotes the running coupling constant that satisfies the NNLO
 renormalization group equation:
\begin{equation}
\mu \frac{da}{d\mu} = - b\;a^{2} [ 1 + c_{1}a + c_{2}a^{2}],
\label{rge}
\end{equation}
 and where
\begin{equation}
r_{1}(\mu/P) = r_{1}^{(0)} + b \ln(\mu/P),
\end{equation}
\begin{equation}
r_{2}(\mu/P) = r_{2}^{(0)} + (c_{1} + 2r_{1}^{(0)}) b \ln(\mu/P) +
 (b \ln(\mu/P))^{2},
\end{equation}
 The coefficients  $r_{i}^{(0)}$ are
 independent of $\mu$. The commonly used renormalization group
 improved expression for $R$ is obtained by assuming that
 $\mu=k P$, where $k$ is a constant.
 In the following it is assumed, that the quark mass effects are
 approximated by a step-function energy dependence of the number
 of active quark flavors.

 The results of perturbative calculations are usually expressed
 in the modified minimal subtraction ($\overline {MS}$) scheme
 \cite {msb}. Other choices of RS are possible, corresponding to
 different choices of the finite parts of renormalization
 constants. Results in various schemes are related by a finite
 renormalization, which in our approximation is equivalent to
 a redefinition of the coupling constant:
\begin{equation}
a_{\overline {MS}}(\mu) = a(\mu)[ 1 + A_{1}\,a(\mu) +
 A_{2}\,a^{2}(\mu)].
\end{equation}
 The constants $A_{1}$ and $A_{2}$ may in principle be arbitrary.
 The expansion coefficients $r_{1}$, $r_{2}$ and $c_{2}$ depend
 on the choice of RS. Also the parameter $\Lambda$ does depend
 on the RS \cite{CelGon}. The relevant formulas describing this
 dependence are collected for example in \cite{R92}. However, in
 NNLO there exists an RS independent combination of the
 expansion coefficients \cite{PMS,Dhar,DharGupta}:
\begin{equation}
\rho_{2}=c_{2}+r_{2}-c_{1}r_{1}-r_{1}^{2}.
\label{inv}
\end{equation}
 As was pointed out in \cite{R92}, a convenient description of the
 RS dependence of $R^{(2)}$ is obtained when $r_{1}$ and $c_{2}$
 are chosen as independent parameters characterizing the
 choice of RS. The parameter $r_{1}$ controls the degree of
 freedom in the approximants which is equivalent to the
 freedom of choice of the renormalization scale $\mu$, and
 $c_{2}$ controls -- in the terminology of
 Stevenson \cite{PMS} --
  the dependence on the choice of the renormalization
 convention. With this parametrization the NNLO RG-improved
 expression for $R$ may  be written in the form \cite{R92}:
\begin{equation}
R^{(2)}(P/\Lambda_{\overline{MS}};r_{1},c_{2})=
a(P)[1+r_{1}\,a(P)+r_{2}(r_{1},c_{2})\,a^{2}(P)],
\label{impr}
\end{equation}
 where $r_{2}$ is expressed as a function of $r_{1}$ and $c_{2}$
 using the Eq.~\ref{inv}. The numerical value of the running
 coupling constant is obtained from the implicit equation:
\begin{equation}
b \ln(\frac{P}{\Lambda_{\overline{MS}}})=
r_{1}^{\overline {MS}}-r_{1}+\Phi^{(2)}(a,c_{2}),
\label{intrge}
\end{equation}
 where
\begin{equation}
\Phi^{(2)}(a,c_{2})=c_{1} \ln (\frac {b}{2c_{1}})+\frac {1}{a} +
 c_{1} \ln (c_{1}a)+O(a),
\end{equation}
 which is obtained by integrating the Eq.~\ref{rge} with an
 appropriate boundary condition.
 The explicit form of $\Phi(a,c_{2})$ may be found for example in
 \cite{rge}. The presence of $\Lambda_{\overline {MS}}$ in the
 Eq.~\ref{intrge} is a result of taking explicitly into account
 the one loop Celmaster-Gonsalves relation between
 $\Lambda$ parameters
 in different schemes \cite{CelGon}, which is valid to all orders of
 perturbation expansion.

The fact,  that numerical value of the predictions depends on the choice
 of RS,  stimulated the interest in prescriptions, which
 pick up schemes distinguished by some arguments of physical or
 mathematical character. Several such prescriptions for selecting
 a ``good'' RS have been discussed, like the prinicple
 of Fastest Apparent Convergence (FAC) \cite{GrunFAC}, the
 Principle of Minimal Sensitivity (PMS) \cite{PMS}, the
 momentum subtraction  prescription \cite{CelSiv,DharGuptaMOM}, and
 other \cite{BLM}. However, regardless
 of our choice of the optimal scheme, there is always a continuum
 of schemes which are close to the one
 preferred by us. Since any choice of a preferred RS may have
 only a heuristic motivation, the predictions obtained in
 schemes close to the optimal scheme also must be somehow taken into
 account. A natural way to accomplish this is
 to supplement the prediction in a preferred scheme
 by the estimate of the strength of the scheme dependence.

 In order to obtain a meaningful estimate of the strength of RS
 dependence one should calculate variation of predictions over
 all {\em a priori} acceptable schemes. To this end
 we must provide some well motivated criteria for selecting the
 admissible scheme. The use of some selection criteria cannot be
 avoided, because by taking variation of predictions over
 {\em all} possible
 schemes we would obtain a meaningless result. The reason for this is
 obvious -- clearly one may make a bad choice of scheme, in which the
 perturbation expansion could have unnaturally large expansion
 coefficients, and one may artificially obtain very large higher
 order corrections and very strong RS dependence, without any
 physical significance. As was pointed out in \cite{R92},
 in order to eliminate unnatural schemes
 one may use the expression for the
 RS invariant $\rho_{2}$. This invariant appears as an NNLO expansion
 coefficient in the manifestly RS-invariant evolution equation
 for a physical quantity \cite{Dhar,DharGupta,GrunEC},
 and therefore it may be taken
 as a natural RS-independent characterization of the magnitude
 of the NNLO corrections. The observation made in \cite{R92} was
 that one should avoid using schemes that introduce large
 cancellations in the expression for $\rho_{2}$.

 To make this condition more concrete let us introduce a function
 which describes the degree of
 cancellation in the expression for $\rho_{2}$ in a quantitative way:
\begin{equation}
\sigma_{2}(r_{1},r_{2},c_{2})=|c_{2}|+|r_{2}|+c_{1}|r_{1}|+r_{1}^{2}.
\end{equation}
 The requirement,  that the contributions to $\rho_{2}$ should not
 be too large compared to the magnitude of $\rho_{2}$ itself, may
 be now written in the form:
\begin{equation}
\sigma_{2}(r_{1},r_{2},c_{2}) \leq l\,|\rho_{2}|,
\label{constraint}
\end{equation}
 where $l\geq 1$ is some constant . The constant $l$ determines how strong
 cancellations in the expression for $\rho_{2}$ we want to allow.
 For arbitrary $l$ it is straightforward to obtain an explicit
 description of the region of allowed parameters in the
 $(r_{1},c_{2})$ plane. In the case $\rho_{2}>c_{1}^{2}/4$, which
 is what will be needed for the Bjorken sum rule, let us define:
\begin{eqnarray}
r_{1}^{min}= - \sqrt{\rho_{2}(l-1)/2},\\
r_{1}^{max}= [-c_{1}+\sqrt{c_{1}^{2}+2(l-1)\rho_{2}}\;]/2,\\
c_{2}^{min}= - \rho_{2}(l-1)/2,\\
c_{2}^{max}= \rho_{2}(l+1)/2,\\
c_{2}^{int}= c_{1}r_{1}^{min}+c_{2}^{max}.
\end{eqnarray}
For $c_{2}>0$ the region of allowed parameters is bounded from
 above by the line joining the points $(r_{1}^{min},0)$,
 $(r_{1}^{min},c_{2}^{int})$, $(0,c_{2}^{max})$,
 $(r_{1}^{max},c_{2}^{max})$, $(r_{1}^{max},0)$.
 For $c_{2}<0$ the region of allowed parameters is bounded from
 below by the lines:
\begin{eqnarray}
c_{2}(r_{1})=r_{1}^{2}+c_{2}^{min}\;\;\;\;
for\;\;\;\;  r_{1}^{min}\leq r_{1}\leq 0,\\
c_{2}(r_{1})=r_{1}^{2}+c_{1}r_{1}+c_{2}^{min}\;\;\;\;
 for\;\;\;\; 0\leq r_{1}\leq r_{1}^{max}.
\end{eqnarray}
It is interesting that the FAC scheme, corresponding to
 $r_{1}=0$, $c_{2}=\rho_{2}$, always belongs to the allowed region.
 We also see, that for $\rho_{2}=0$ -- i.e. null NNLO correction --
  the only scheme satisfying
 Eq.~\ref{constraint} is $r_{1}=0$, $c_{2}=0$ (and of course
 $r_{2}=0$).

 For a final specification of the allowed region in the $(r_{1},c_{2})$
 plane we have to choose some value of $l$ in the
 Eq.~\ref{constraint}. To make a meaningful choice let us observe,
 that it is a rather natural requirement to have the PMS scheme
 in the allowed region. The PMS scheme parameters are determined by the
 condition:
\begin{equation}
\frac {\partial R^{(2)}}{\partial r_{1}} = 0,\;\;\;\;\;
\frac  {\partial R^{(2)}}{\partial c_{2}} = 0.
\end{equation}
An approximate solution of these quations has the form
\cite{PennWrig}:
\begin{equation}
r_{1}^{PMS}=0(a^{PMS}),\;\;\;\;\;\;\;\;
c_{2}^{PMS}=\frac {3}{2} \rho_{2} + 0(a^{PMS}).
\label{aPMS}
\end{equation}
It is easy to see that taking $l=2$ we would have the PMS scheme
 right at the boundary of the allowed region in the
 $(r_{1},c_{2})$ plane.

Therefore, in order to have an estimate of the strength of the
 scheme dependence in NNLO we propose to calculate the variation
 of predictions using Eq.~\ref{impr} and Eq.~\ref{intrge},
 when the scheme
 parameters $r_{1}$ and $c_{2}$ are changed within the region
 satisfying the Eq.~\ref{constraint} with $l=2$.

 Let us now consider the QCD correction $R_{Bj}$ to the Bjorken
 sum rule for the polarized structure functions in the
 deep-inelastic scattering \cite{Bjork}:
\begin{equation}
\int_{0}^{1}\;dx\;[g_{1}^{p}(x,Q^{2})-g_{1}^{n}(x,Q^{2})]=
\frac{1}{6} \frac{g_{A}}{g_{V}}[1-R_{Bj}(Q^{2})].
\end{equation}
This sum rule has been recently a subject of intense experimental
 \cite{exp,combined} and theoretical \cite{3loop,pheno}
 investigation. The QCD perturbation expansion
 for $R_{Bj}$ has the form of Eq.~\ref{R}, with $P^{2}=Q^{2}$,
 where  the expansion
 coefficients in the $\overline{MS}$ scheme with four quark
 flavors are given by \cite{1or2loop,3loop}:
\begin{equation}
r_{1}^{\overline{MS}}=3.25\;\;\;\;\;\;\;\;
r_{2}^{\overline{MS}}=13.8503.
\end{equation}
Together with $c_{1}=1.54$, $c_{2}^{\overline{MS}}=3.0476$ this
 implies $\rho_{2}=1.3304$ ($b=25/6$). In Figure 1 we show
 $R_{Bj}^{(2)}$ as a function of $r_{1}$ and $c_{2}$ for
 $P/\Lambda_{\overline{MS}}=8.6$ (this corresponds to
 $<Q^{2}>=5\,GeV^{2}$ -- as used by the SMC Collaboration to present
 the combined experimental data \cite{combined} -- and
 $\Lambda_{\overline{MS}}^{(4)}=0.26\,GeV$ as preferred by the
 Particle Data Group \cite{PDG}), together with the region of
 allowed values for the scheme parameters. We see that $R_{Bj}^{(2)}$
 has indeed a saddle point, representing prediction in the
 PMS scheme, close to the $l=2$ allowed region. We have:
\begin{equation}
R_{Bj}^{PMS}(8.6)=R_{Bj}^{(2)}(8.6;0.0852,2.1302)=0.126525.
\end{equation}
 The estimate
 of the strength of the scheme dependence is obtained by comparing the
 predictions inside the allowed region. We find that the extremal
 values are attained at the boundary of the allowed region:
\begin{eqnarray}
R_{Bj}^{max}(8.6)=R_{Bj}^{(2)}(8.6;-0.668,-0.219)=0.12697,\\
R_{Bj}^{min}(8.6)=R_{Bj}^{(2)}(8.6;0.352,0)=0.12597.
\end{eqnarray}
 For comparison we
 also show in Figure~1 the range of scheme parameters
 satisfying the constraint
 given by Eq.~\ref{constraint} with $l=3$, in which case a small
 neighbourhood of the PMS parameters is taken into account. In this
 case we obtain:
\begin{eqnarray}
R_{Bj}^{max}(8.6)=R_{Bj}^{(2)}(8.6;-0.838,-0.628)=0.12722,\\
R_{Bj}^{min}(8.6)=R_{Bj}^{(2)}(8.6;0.617,0)=0.12551.
\end{eqnarray}

In Figure~2 we show the prediction for $R_{Bj}^{(2)}$ as a function
 of $P/\Lambda_{\overline{MS}}$ near the point
 $P/\Lambda_{\overline{MS}}=8.6$.
 This figure shows how the uncertainty of the predictions
 arising from the RS dependence may affect the fit of
 $\Lambda_{\overline{MS}}$ to possible experimental results.
 We see that the obtained scheme dependence appears to be very small.
 In fact it is much smaller than the uncertainty arising from the
 estimated error in the fitted value of $\Lambda_{\overline{MS}}$ --
 the Particle Data Group gives
 $\Lambda_{\overline{MS}}^{(4)}=0.316-0.214\,VeV$, which for
 $Q^{2}=5\;GeV^{2}$ gives variation of $Q^{2}/\Lambda^{2}$ in the
 range $(7.08)^{2}-(10.45)^{2}$. The scheme dependence
 uncertainty is also  much smaller than the experimental errors in
 the presently available experimental data \cite{exp,combined}.
 This shows that
 the $n_{f}=4$ NNLO formula for the QCD corrections to the
 polarized Bjorken sum rule seems to be a potentially good source
 of precise information on $\Lambda_{\overline{MS}}$ once the
 experimental accuracy is improved.

 Let us also note that the $\overline{MS}$ coefficients lie outside
 the allowed region indicated in Figure~1, i.e.\ in our
 approach the $\overline{MS}$ scheme
 has to be considered as a scheme with unnaturally
 large expansion coefficients. Also,
 $R_{Bj}^{\overline{MS}}(8.6)=0.11969$ lies outside the range of
 values obtained in our estimate of scheme dependence uncertainties.
 However, the difference between $R_{Bj}^{\overline{MS}}$ and
 $R_{Bj}^{PMS}$ is still rather small for
 $Q^{2}/\Lambda_{\overline{MS}}^{2}$ close to $(8.6)^{2}$.

 For the sake of comparison it is interesting to consider the
 RS dependence of the NNLO expression for $R_{Bj}$
 for $n_{f}=3$.
 Taking three quark flavors we have:
\begin{equation}
r_{1}^{\overline{MS}}=3.5833\;\;\;\;\;\;\;\;
r_{2}^{\overline{MS}}=20.2153,
\end{equation}
which together with $c_{1}=16/9$ and $c_{2}^{\overline{MS}}=4.471$
 implies $\rho_{2}=5.4757$ ($b=9/2$). This is a significantly larger value
 than for four flavors. The dependence of $R_{Bj}^{(2)}$ on
 $r_{1}$ and $c_{2}$ for $n_{f}=3$ for
 $P/\Lambda_{\overline{MS}}=8.6$ is shown in Figure~3.
 We find:
\begin{equation}
R_{Bj}^{PMS}=R_{Bj}^{(2)}(8.6;0.33,8.83)=0.121634.
\end{equation}
Repeating the procedure used for $n_{f}=4$ we obtain for
 the $l=2$ allowed region:
\begin{eqnarray}
R_{Bj}^{max}(8.6)=R_{Bj}^{(2)}(8.6;-1.655,0)=0.12748,\\
R_{Bj}^{min}=R_{Bj}^{(2)}(0.989,0)=0.11757.
\end{eqnarray}
For the $l=3$ allowed region we find:
\begin{eqnarray}
R_{Bj}^{max}(8.6)=R_{Bj}^{(2)}(8.6;-2.340,0)=0.13412,\\
R_{Bj}^{min}(8.6)=R_{Bj}^{(2)}(8.6;-2.340,6.791)=0.10683.
\end{eqnarray}
Note that also in this case the $\overline{MS}$
 coefficients lie
 outside the allowed range for scheme parameters. We have
 $R_{Bj}^{\overline{MS}}=0.11102$, which is close to the
 lower limit of values obtained by varying over the $l=3$ allowed
 region. We see that the theoretical uncertainty in the
 $n_{f}=3$ formula is larger than in the $n_{f}=4$ formula.
 This is a result of stronger dependence of $R^{(2)}$ on the
 scheme parameters in this case, and of the larger range of
 parameters which may be considered to be natural. Using
 our method we may now quantitatively compare reliability of the
 $n_{f}=3$ and $n_{f}=4$ predictions for $R_{Bj}$. Concerning the
 energy dependence of the theoretical uncertainties in this
 case let us note, that for lower values
 of $Q^{2}/\Lambda^{2}$ the RS dependence of the three-flavor
 expression, estimated according to our prescription, is considerably
 larger than that found above. The problem of the low energy behavior
 of $R_{Bj}^{(2)}$ and a possible method for reducing scheme
 dependence at low energies would be discussed in a
 separate note \cite{R94}.

 We conclude this note with several remarks.

 1. It has to be emphasized that
 our estimate of the strength of the
 RS dependence does not provide an error estimate in the
 mathematical sense, i.e.\ there is no theorem that would
 guarantee that the true result lies within the obtained range of
 variation for the prediction. The method described above provides only
 a consistent framework for making an unbiased educated guess about
 possible theoretical precision of the prediction. The method avoids
 unnecessary constraints on the considered renormalization schemes,
 it eliminates from analysis  the unnatural schemes, and it is sensitive
 to the magnitude of the NNLO correction via dependence on $\rho_{2}$.
 The estimate of the scheme dependence uncertainty obtained in this
 way should be particularly useful for a quantitative comparison
 of reliability of perturbative predictions for different physical
 quantities, evaluated at different energies. In fact one may
 reasonably expect that thus obtained {\em relative} estimate of the
 reliability of perturbative predictions would be insensitive
 to certain freedom in choosing $l$ in the Eq.~\ref{constraint} and
 to possible variations in the choice of the function
 $\sigma_{2}$. The
 proposed method should be also useful in determining the regions of
 applicability of perturbation expansion. Indeed, since we have
 taken great care to eliminate schemes with unnaturally large
 expansion coefficients, in order to avoid introducing any spurious
 RS dependence, then any large variation of predictions obtained
 according to our method would be an unambiguous sign of a limited
 applicability of the NNLO expression in the considered
 energy range.

 2. The RS invariant $\rho_{2}$ plays
 a fundamental role in our
 approach. However, Stevenson in his anlysis of RS dependence
 \cite{PMS} used an expression for the NNLO invariant that differs
 from the one used here by a constant. We think that the expression
 used in this note is more natural for characterization of the
 magnitude of NNLO corrections because it appears  in the RS
 invariant evolution equation for a physical quantity
 \cite{Dhar,DharGupta,GrunEC}.
 To emphasize the RS independent character of the evolution
 equation obtained in \cite{Dhar,DharGupta,GrunEC}
 let us recall the argument indicated
 in \cite{R90}. Let us take $R^{(2)}$ as given by Eq.~\ref{R} in
 {\em arbitrary} scheme, differentiate over $P$, then express
 $a(P)$ in terms of $R^{(2)}$, and finally express
 $P(dR^{(2)}/dP)$ as a series expansion in $R^{(2)}$.
 We obtain:
\begin{equation}
P \frac{dR^{(2)}}{dP}= - b (R^{(2)})^{2} [ 1 + c_{1} R^{(2)}
 + \rho_{2} (R^{(2)})^{2}  +
 \sum_{k=3}^{\infty} {\overline {\rho_{k}}} (R^{(2)})^{k} ].
\end{equation}
The first three expansion coefficients in this equation are
 RS independent and coincide with the coefficients obtained
 in \cite{Dhar,DharGupta,GrunEC}.
 Higher order expansion coefficients in this
 equation are scheme dependent, which is a result of the fact
 that we have only used NNLO approximation for a physical
 quantity. (If higher order approximant would be taken, then
 more coefficients in the above equation would become
 independent of the choice of RS.) This shows, that the equation
 obtained in \cite{Dhar,DharGupta,GrunEC}
 is indeed RS independent, and that it is  not a result
 of a particular choice of the scheme.

 Another argument supporting our choice is the fact that
 $\rho_{2}$ naturally appears in the leading order solution
 of the PMS equations (Eq.~\ref{aPMS}).

 3. The fact that we base our
 method on the existence of the
 RS invariant $\rho_{2}$ has an obvious consequence, that it
 cannot be applied to quantities for which only next-to-leading
 (NLO) corrections are known. This should not be surprising
 however, because the NLO approximation has a rather special
 status in QCD. In fact, the NLO correction allows only to
 set the proper energy scale for the prediction. For example,
 using the FAC scheme we obtain in NLO an {\em identical} energy
 dependence for {\em all} possible physical observables depending
 on one energy variable, except for a shift on a logarithmic
 energy scale. It is only at NNLO that physical quantities
 receive process-specific corrections. It is therefore natural
 that only at NNLO and beyond one may test reliability of the
 perturbation expansion. This indicates a fundamental
 importance of the NNLO calculations for a proper
 comparison of theoretical
 predictions with experimental data. Let us also note that it is
 straightforward to extend our approach to the case of four-loop
 or even higher order corrections if such calculations would
 ever be done.

 4. Let us discuss some other methods
 of estimating the reliability of
 perturbative predictions. A simplest alternative is to compare
 NLO and NNLO predictions, taking the difference as an estimate
 of the theoretical uncertainty. Unfortunately, it is obvious that
 such an estimate would be strongly dependent on the choice of
 scheme. One may also try to guess the magnitude of higher order
 corrections. Again, the problem here is the RS dependence of such
 an approach. Note also that the asymptotic expression for
 higher order corrections obtained in  \cite{west}, which has been
 sometimes used to evaluate the theoretical
  uncertainty, is known to be erroneous
 \cite{chylafischer}. Finally, in \cite{chylakataev}
 it is proposed to obtain theoretical
 uncertainty by comparing predictions obtained in the PMS, FAC and
 $\overline {MS}$ schemes. Unfortunately, this method also may
 give misleading results, since the predictions in the PMS and
 FAC approach are known to lie rather close to each other regardless of the
 magnitude of the NNLO correction, and the choice of $\overline {MS}$
 in such procedure is completely arbitrary, without
 any theoretical or phenomenological
 motivation. In fact, as was shown above, comparing PMS and
 $\overline {MS}$ we may obtain {\em larger} uncertainty than
 that suggested by our method.

 5. The method described in this
 note has been applied to other
 quantities for which the NNLO expansion is known
 \cite{R94,RSa,RSb}.

 The results
 for the Gross-Llewellyn-Smith sum rule are similar to those
 obtained above for the Bjorken sum rule. The NNLO expression
 for $n_{f}=4$ appears to have small scheme uncertainty, while
 the $N_{f}=3$ expression seems to be less precise, particularly
 at lower energies.

 In the case of the QCD corrections to the
 tau lepton decay we
 confirm the observations made in \cite{R92}.
 By varying predictions over the
 set of schemes allowed by the constraint given  by
 Eq.~\ref{constraint} we find that the frequently used NNLO
 perturbative expression is unstable against the change of RS.
 (In \cite{R92} the condition on the allowed schemes was not
 specified in a quantitative way.) However, if one takes a
 contour integral representation for $R_{\tau}$,
 one obtains
 a stable result \cite{RSa}, with  theoretical uncertainty of the same
 order as the present experimental uncertainty.
 Neglecting nonperturbative contributions for simplicity and
 fitting $\Lambda_{\overline{MS}}$ to the experimental central
 value $R_{\tau}^{exp}=0.2$ we obtain $0.396\,GeV$ in the
 generalized PMS method, and $0.386-0.415\,GeV$ variation for the
 $l=2$ allowed region for the scheme parameters.

 Application of our method to the case of the $n_{f}=5$
 QCD corrections to the $e^{+}e^{-}$ annihilation into
 hadrons reveals a surprisingly large RS dependence, despite rather
 high characteristic energy scale. The large scheme uncertainty
 is a consequence of a large value of the RS invariant
 $\rho_{2}$. However, a closer analysis shows that large value
 of this invariant is due mainly to the effect of analytic
 continuation from the spacelike to timelike momenta. A proper
 treatment of the analytic continuation results in a
 considerable improvement of the stability of the predictions
 with respect to change of the RS \cite{RSb}. Similar remarks apply to
 the case of the QCD corrections to the $Z$-boson decay rate
 to hadrons, which is however more complicated because of
 the presence of mass dependent corrections.

 Summarizing, we may say, that a method has been proposed for
 analyzing the uncertainties in the NNLO predictions, introduced
 by arbitrariness in the choice of the renormalization scheme. This
 method is based on a condition that eliminates schemes which
 have expansion coefficients giving unnaturally large cancellations
 in the expression for the characteristic NNLO RS invariant.
 Application of the method has been illustrated using as an
 example the QCD corrections to the Bjorken sum rule for
 polarized structure functions. It was found that for four quark
 flavors these corrections have small scheme dependence uncertainty.
 Our considerations show the fundamental importance of the
 NNLO corrections for a proper evaluation of the perturbative
 QCD predictions.

 Author is grateful to the High Energy Physics
 group at the Argonne National Laboratory for hospitality and
 discussion.

\newpage
\section*{Figure Captions}
\noindent Fig. 1. Contour plot of the NNLO prediction for
 $R_{Bj}$ as a function of $r_{1}$ and $c_{2}$, for $n_{f}=4$
 and $Q^{2}/\Lambda_{\overline{MS}}^{2}=(8.6)^2$. The region
 of values of the scheme parameters satisfying the constraint
 given by Eq.~\ref{constraint} is indicated for $l=2$ (smaller
 region) and for $l=3$. The dashed line corresponds to
 $R_{Bj}^{(2)}=0.126526$.

\noindent Fig. 2. The NNLO prediction for $R_{Bj}$ as a function
 of $P/\Lambda_{\overline{MS}}$ ($P^{2}=Q^{2}$). The thick broken
 line in the middle denotes the PMS prediction. The thin solid
 lines denote the variation of the predictions when $r_{1}$ and
 $c_{2}$ are changed within a region defined by Eq.~\ref{constraint}
 with $l=2$. Thin broken lines correspond to $l=3$.

\noindent Fig. 3. Same as in Fig.\ 1  but for $n_{f}=3$. The dashed
 line corresponds to $R_{Bj}^{(2)}=0.121635$.
\end{document}